\documentclass[epj]{webofc}
\usepackage[utf8]{inputenc}
\usepackage[varg]{txfonts}   
\usepackage{booktabs}
\usepackage{xcolor}
\definecolor{darkred}{rgb}{0.4,0.0,0.0}
\definecolor{darkgreen}{rgb}{0.0,0.4,0.0}
\definecolor{darkblue}{rgb}{0.0,0.0,0.4}
\usepackage[bookmarks,linktocpage,colorlinks,
    linkcolor = darkred,
    urlcolor  = darkblue,
    citecolor = darkgreen]{hyperref}
\usepackage{subfigure}
\wocname{EPJ Web of Conferences}
\woctitle{Lattice2017}
\begin{document}
%
\selectlanguage{english}
\title{%
Lattice calculation of hadronic tensor of the nucleon
}
\author{%
\firstname{Jian} \lastname{Liang}\inst{1} \and
\firstname{Keh-Fei} \lastname{Liu}\inst{1} \and
\firstname{Yi-Bo}  \lastname{Yang}\inst{2}
}
\institute{%
Department of Physics and Astronomy, University of Kentucky, Lexington, KY 40506, USA
\and
Department of Physics and Astronomy, Michigan State University, East Lansing, MI 48824, USA
}
\abstract{We report an attempt to calculate the deep inelastic scattering structure functions from the hadronic tensor calculated on the
lattice. We used the Backus-Gilbert reconstruction method to address the inverse Laplace transformation for the analytic continuation
from the Euclidean to the Minkowski space.
}
\maketitle

\section{Hadronic tensor on the lattice}\label{eht}
\hspace{0.5cm}
     There has been a lot of interest and developments in calculating the structure functions and parton distribution functions on the lattice 
in recent years~\cite{Liu:2016djw}. The Euclidean hadronic tensor has been formulated in the path-integral formalism~\cite{Liu:1993cv,Liu:1998um,Liu:1999ak}. The hadronic tensor in deep inelastic scattering, from which the parton distribution functions are obtained through the factorization theorem, can be calculated from its Euclidean counterpart via an inverse Laplace transform~\cite{Liu:1993cv,Liu:1999ak,Liu:2016djw}. It has been revealed that, in addition to the valence partons, there are two types of sea partons -- connected sea (CS) and disconnected sea (DS) in three topologically distinct path-integral diagrams. The extended evolution equations to accommodate both the connected and disconnected sea partons are derived~\cite{Liu:2017lpe}. It is essential to have separately evolved CS and DS partons so that comparison with lattice calculations of unpolarized and polarized moments of PDF can be made. Only with the extended evolution equations will the CS and DS partons remain separated at different $Q^2$ to facilitate global fitting of PDF with separated CS and DS partons.

The definition of hadronic tensor in the Minkowski space is
\begin{eqnarray}   \label{W_Min}
W_{\mu\nu}(q^2,\nu)\!\! &=&\!\! \frac{1}{4\pi}\int d^4z e^{iq\cdot z}\langle p|J_\mu^\dagger(z)J_\nu(0)|p \rangle_{\rm{spin\,\, ave.}} \nonumber \\
\!\! &=&\!\! \frac{1}{2} \sum_n \int \prod_{i =1}^n \left[\frac{d^3 p_i}{(2\pi)^3 2E_{pi}}\right]  \langle N|J_{\mu}(0)|n\rangle
\langle n|J_{\nu}(0) | N\rangle_{\rm{spin\,\, ave.}}(2\pi)^3 \delta^4 (p_n - p - q) . 
\end{eqnarray}
where $|p\rangle$ is the nucleon state and $J_\mu$ is the vector current. 
In the Euclidean path integral formalism, the hadronic tensor related nucleon matrix element can be expressed as
the ratio of the four-point function and the two-point function.
To be specific, they are
\begin{equation}
C_4 (\vec{p}, \vec{q}, \tau)=\sum_{\vec{x}_f}e^{-i\vec{p}\cdot\vec{x}_f}\sum_{\vec{x}_2\vec{x}_1}e^{-i\vec{q}\cdot(\vec{x}_2-\vec{x}_1)}
\langle \chi_N(\vec{x}_f,t_f) J_\mu(\vec{x}_2,t_2) J_\nu(\vec{x}_1,t_1) \bar{\chi}_N(\vec{0},t_0) \rangle,
\end{equation}
\begin{equation}
C_2 (\vec{p}, \tau) =\sum_{\vec{x}_f}e^{-i\vec{p}\cdot\vec{x}_f}
\langle \chi_N(\vec{x}_f,t_f) \bar{\chi}_N(\vec{0},t_0) \rangle,
\end{equation}
where $\chi_N$ is the nucleon operator and $\tau=t_2-t_1$. The $\tau$ dependent Euclidean hadronic tensor is
\begin{equation}
\tilde{W}(\vec{p},\vec{q},\tau)\overset{t_f\gg t_2, t_1\gg t_0}{=}\frac{E_N{\rm Tr}[\Gamma_eC_4 (\vec{p}, \vec{q}, \tau)]}
{m_N{\rm Tr}[\Gamma_eC_2(\vec{p}, \tau) ]},
\end{equation}
where $E_N$ and $m_N$ are the energy and mass of the nucleon and $\Gamma_e=\frac{1+\gamma_4}{2}$
is the unpolarized spin projector.
After inserting the complete set of intermediate states, we have
\begin{equation}  \label{W_tau}
\tilde{W}_{\mu\nu}(q^2,\tau) = \frac{1}{4 \pi}\sum_n \left(\frac{2 m_N}{2 E_n}\right) \delta_{\vec{p}+\vec{q}, \vec{p_n}}
\langle p|J_\mu|n\rangle\langle n|J_\nu|p\rangle_ {\rm{spin\,\, ave.}}e^{-(E_n-E_N)\tau}.
\end{equation}
We see that in Eq.~(\ref{W_tau}) there is an exponential dependence on the Euclidean $\tau$. It will exponentially decay when the 
lowest $E_n$ is heavier than $E_N$ and it will exponentially grow when the lowest $E_n$ is lighter than $E_N$.
To analytic continue to the Minkowski space, an inverse Laplace transform is needed, i.e. 
\begin{equation}  \label{wmunu} 
W_{\mu\nu}(q^2,\nu) = \frac{1}{2m_Ni} \int_{c-i \infty}^{c+i \infty} d\tau\,
e^{\nu\tau} \widetilde{W}_{\mu\nu}(\vec{q}^{\,2}, \tau),
\end{equation} 
with $c > 0$. This is basically doing the anti-Wick rotation back to the Minkowski space to recover the delta function in energy as shown
in Eq.~(\ref{W_Min}).

\begin{figure}[bph]
   \centering
   \subfigure[valence and connected sea parton $q$(V+CS)]%
             {\includegraphics[width=0.3\textwidth,clip]{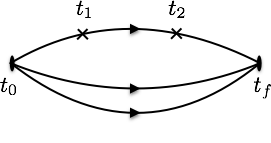}}\hfill
   \subfigure[connected sea anti-parton $\bar{q}$(CS)]%
             {\includegraphics[width=0.28\textwidth,clip]{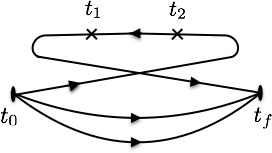}}\hfill
   \subfigure[disconnected sea parton and anti-parton $q$(DS) and $\bar{q}$(DS)]%
             {\includegraphics[width=0.26\textwidth,clip]{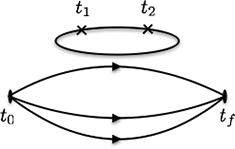}}\hfill
   \subfigure[suppressed by $O(1/Q^2)$]%
             {\includegraphics[width=0.3\textwidth,clip]{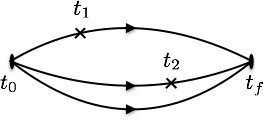}}
   \subfigure[suppressed by $O(1/Q^2)$]%
             {\includegraphics[width=0.3\textwidth,clip]{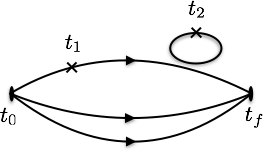}}\hfill
   \caption{Topologically distinct diagrams in the Euclidean-path integral formulation.}
   \label{insertions}
\end{figure}

The topologically distinct insertions are shown in Figure~\ref{insertions}. The first three involve leading-twist contributions
from the valence +CS partons in (a), the CS antipartons in (b), and the DS partons in (c). 
The last two insertions ((d) and (e)) are higher-twist contributions which are suppressed by $O(1/Q^2)$ and will be ignored.

We will focus on the first two insertions in this work. It has been pointed out that 
the second insertion are the connected sea $\bar{u}$ and $\bar{d}$ contribution, 
which are responsible for the Gottfried sum rule violation \cite{Liu:1993cv,Liu:1999ak}. 

\bigskip

It is noted that this approach has the advantage that no renormalization is needed when the conserved vector current is
used and only finite renormalization is needed for local vector current. Furthermore, the structure functions from the hadronic tensor are frame independent so that they can be calculated with the nucleon at rest or at low lattice momenta. However, the inverse Laplace transform, which is needed to convert the hadronic tensor from the Euclidean space to Minkowski space, is a challenge. We shall test a numerical approach to address the inverse problem.

\section{Backus-Gilbert reconstruction}\label{bg}
\hspace{0.5cm}
The most challenging part of our calculation is to convert the Euclidean hadronic tensor $\tilde{W}(q^2,\tau)$ as a function in $\tau$
to the Minkowski ${W}(q^2,\nu)$ in $\nu$ through certain numerical approximation. 
This is a typical case of the inverse problem. We shall consider the Backus-Gilbert reconstruction method as suggested 
in~\cite{Hansen:2017mnd}.

A general form of the inverse problem is
\begin{equation}
c_i = \int k_i(x)w(x)dx,
\end{equation}
where $c_i$ are the discrete data points that we have, $k_i(x)$ are the corresponding integral kernels and $w(x)$ is the desired continuum function. In principle, one cannot get a unique solution for $w(x)$ since the number of input data points is always finite.
In other words, we do not have all the information needed to reconstruct every detail of the continuum function.
Even though the continuum function $w(x)$ can be discretized into $w_j$, the number of points of $w_j$ we need will in general be larger than that of $c_i$. So the inverse problem is, in principle, an ill-defined problem. 
However, we are still able to find the most probable solution to the problem by using some algorithms, such as the maximum 
entropy method \cite{Asakawa:2000tr} and the Backus-Gilbert method \cite{Backus1,Backus2}.

In our case, the input data are $\tau$-dependent Euclidean correlation function, the kernel is the Laplace kernel and the function $\omega({\nu})$ 
is the spectral density 
\begin{equation}
C(\tau) = \int e^{-\omega(\nu)\tau}\,\omega(\nu)\,d\nu.
\end{equation}

Another reason to use the Backus-Gilbert method is when considering the exclusive multi-hadron final states  on the lattice,
the finite volume effects must be handled correctly. Based on \cite{Hansen:2017mnd}, the regularized delta-function introduced in 
the Backus-Gilbert method will ensure that the reconstructed spectrum density function has a well-defined infinite volume limit.

\subsection{Numerical details}\label{bg-1}
\hspace{0.5cm}
The central idea of the Backus-Gilbert method is that a linear combination of the kernel functions can be used to approximate the delta function,
\begin{equation}
\label{r-delta}
\sum_t e^{-\omega(\nu)t} a_t(\nu_0) \sim \delta(\nu-\nu_0).
\end{equation}
Assuming the equation holds, we will have
\begin{equation}
\sum_t a_t(\nu_0)C(t) = \int \delta(\nu-\nu_0)\omega(\nu)d\nu=\omega(\nu_0).
\end{equation}
If we can compute $\vec{a}(\nu_0)$ ($\vec{a}$ means the coefficients vector $(a_0, a_2, ...)^T$) 
for each $\nu_0$ and for each $\nu_0$ Eq.~(\ref{r-delta}) always holds, then we can have the value of  the function $\omega(\nu)$ at
arbitrary $\nu$, which means that the inverse problem is solved. Practically, the finite number of kernel functions cannot span a complete function base and, therefore,
the delta function approximated form the linear combination cannot be a true delta function, but a broadened Gaussian-like function instead (the regularized delta function).
The more data points we have, the more sharp the Gaussian-like function will be. 
The sharpness of the Gaussian-like function actually embodies the resolution of the reconstruction.
This is consistent with our intuition,  the more data points we have, the more information we can extract.

On the other hand, to solve the coefficients $\vec{a}(\nu_0)$, we can first define a function $K(\nu_0, \vec{a})$ of $\vec{a}$ as the ``deltaness'' of the Gaussian-like function,
\begin{equation}
K(\nu_0, \vec{a})=12\int(\nu-\nu_0)^2\left[ \sum_t e^{-\omega(\nu)t} a_t -\delta(\nu-\nu_0) \right]^2d\nu,
\end{equation}
and the coefficients $\vec{a}$ which minimize the value of $K$ will serve as the desired $\vec{a}(\nu_0)$. Therefore to solve for the coefficients $\vec{a}(\nu_0)$ we need to solve the
linear equations 
\begin{equation}
\frac{\partial K(\nu_0, \vec{a})}{\partial a_t} = 0.
\end{equation}
Moreover, to remove the trivial solution of $\vec{a}=\vec{0}$, a Lagrange multiplier is added into the equations.
The error of the Backus-Gilbert result $\sigma(\nu_0)$ is defined as
\begin{equation}   \label{sigma}
\sigma^2(\nu_0) = \vec{a}^T({\rm Cov})\vec{a},
\end{equation}
where $({\rm Cov})$ is the covariance matrix of the input data.
Since in general the errors of the data are not negligible, the very details of the data are not accurate (altered randomly by the noise),
one needs to introduce an SVD cutoff to remove these imprecise small eigenvalues when solving the linear equations.
This SVD cutoff is the only tunable parameter when using the Backus-Gilbert method. When the cutoff is set to be small, 
more eigenvalues will be kept and thus better resolution one can have. However, the small eigenvalues above the cut 
will result in larger $a_t$ values and, consequently, larger final errors from Eq.~(\ref{sigma}).
On the other hand, when the cutoff is set to be larger, less eigenvalues will be kept and thus the resolution will be worse, but the errors will be smaller and the results will be more stable. So this leads to a trade-off. In practice, the SVD cutoff is chosen according to the errors of our data. We try to find a stable window of the cutoff which gives consistent results and then choose the one producing the smallest errors in that window. 

\subsection{Toy model tests}\label{bg-2}
\begin{figure}[tph]
   \centering
   \subfigure[Discrete spectrum result.]%
             {\includegraphics[width=0.475\textwidth,clip]{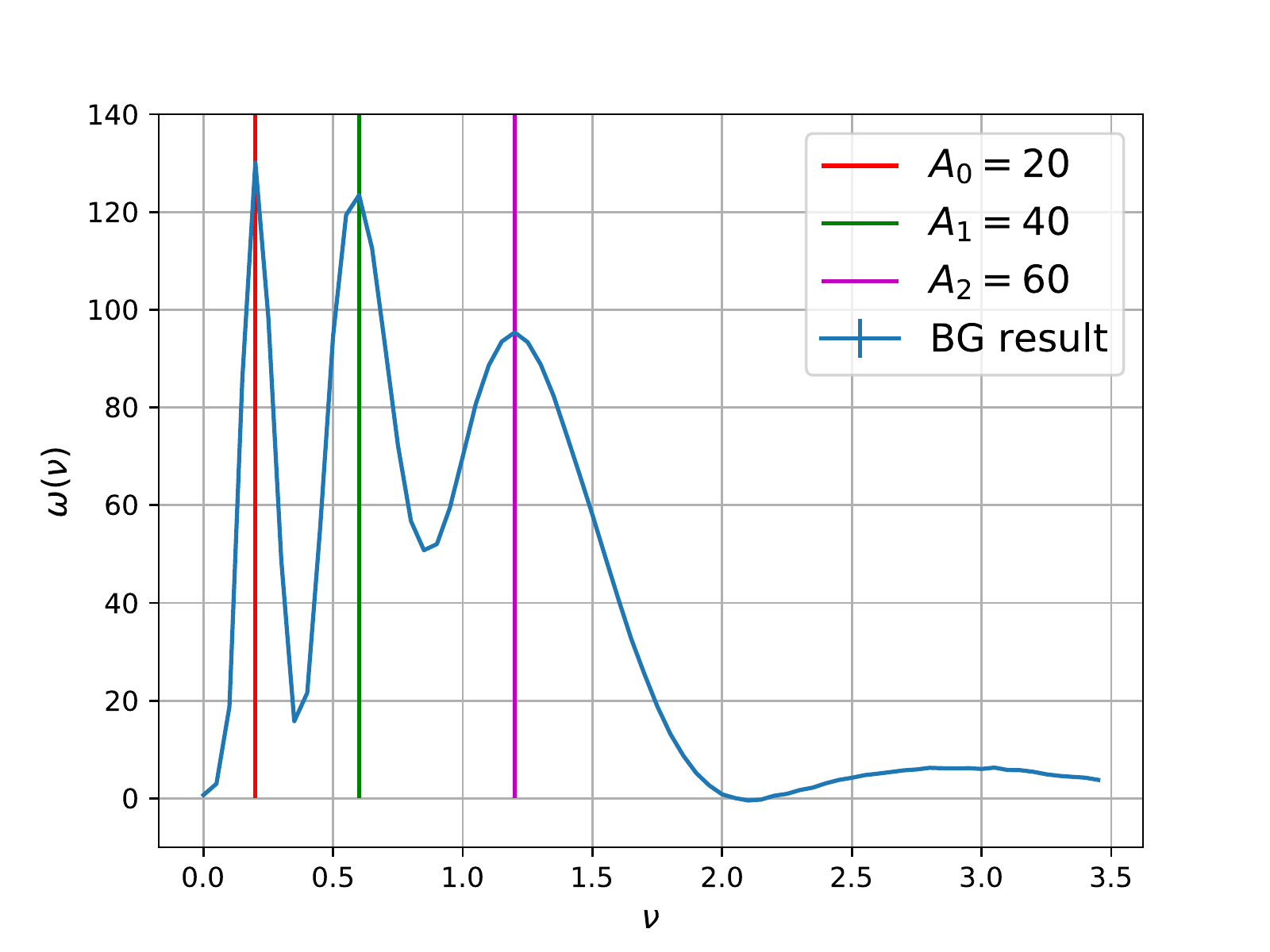}}\hfill
   \subfigure[Discrete spectrum plus dense (continuum) spectrum result.]%
             {\includegraphics[width=0.475\textwidth,clip]{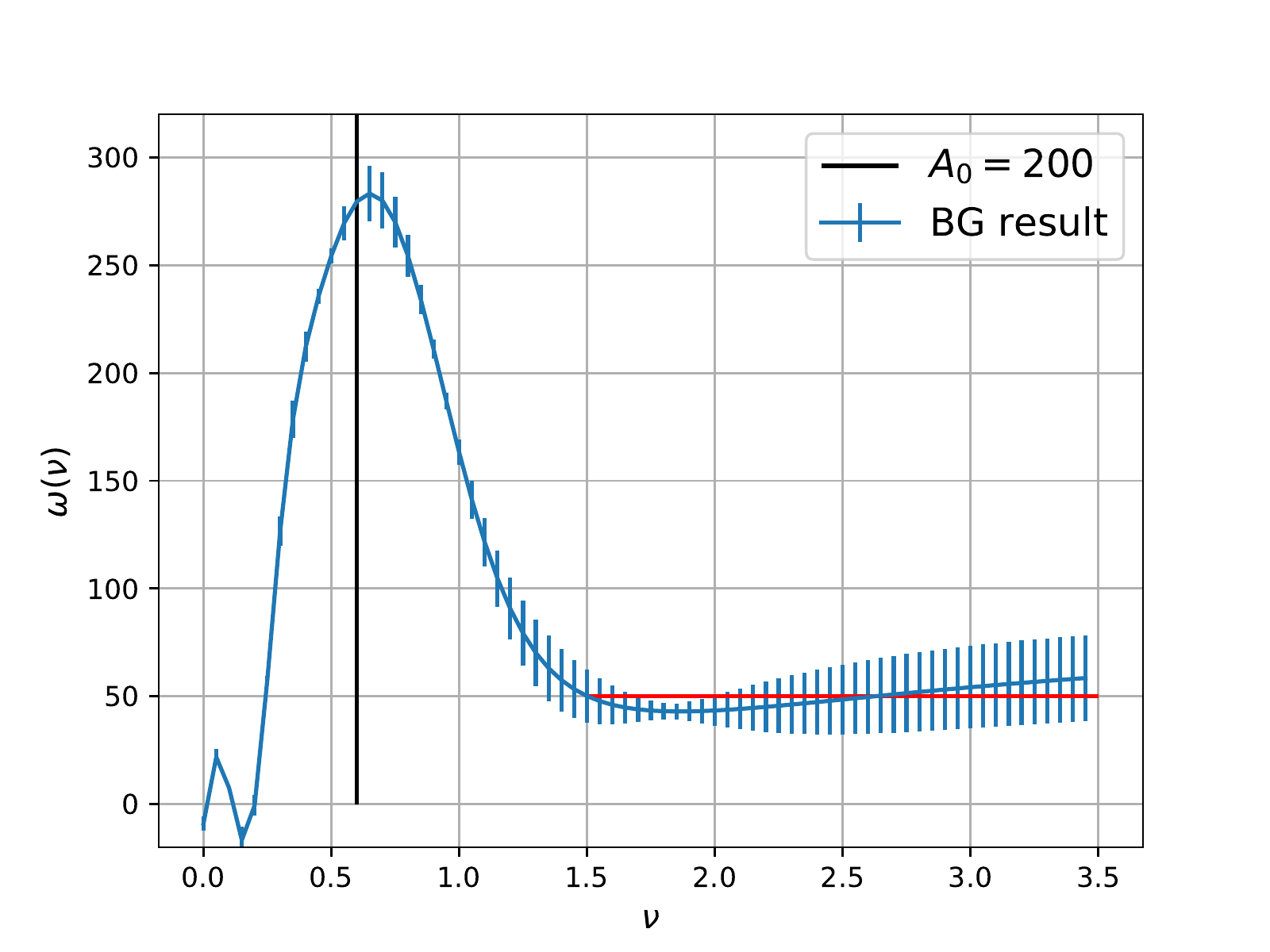}}
   \caption{Backus-Gilbert reconstruction from toy model data. Cyan lines are the resultant spectrum density functions, 
   vertical lines representing delta functions show the discrete $\nu$ positions set in the toy models and 
   the horizon line indicates the spectrum density of the input dense spectrum.}
   \label{fig-1}
\end{figure}

\hspace{0.5cm}
To test the Backus-Gilbert method,
first, we generate a mock two-point correlation function containing three mass terms without noise by
\begin{equation}
C^m_2(t) = \sum_{i=0 }^2A_i e^{-m_it}.
\end{equation}
We set $A_i=20, 40$ and $60$ and $m_i=0.2, 0.6$ and 1.2 respectively.
The number of time slices used is 25, which is close to that of the preliminary lattice calculation in the following section.
The Backus-Gilbert reconstructed results are shown in the left panel of Figure~\ref{fig-1}. 
We can see the three peaks of the spectral density function are indeed located close to the input positions of $m_i$ (vertical lines in the plot),
which means the Backus-Gilbert method extracts the correct spectral information from the toy two-point function.
We can also estimate the spectral weight $A_i$ from the spectral function.
To do so, we use Gaussian functions to approximate the peaks.
We know that the integrals of Gaussian functions $A'_ie^{-\frac{(x-\mu_i)^2}{2\sigma_i^2}}$ are 
$I^G_i=A'_i\sqrt{2\pi}\,\sigma_i$, where $A'_i$ are the peak values of the Gaussian functions and $\sigma_i$ are the standard deviations.
Whereas in the ideal case, the three peaks should be delta functions and the integrals give just the input $A_i$ values: $I^d_i=A_i$.
If we assume $I^G_i = I^d_i$ for each peak, we have
\begin{equation}
A_i = A'_i\sqrt{2\pi}\,\sigma_i.
\end{equation}
For convenience sake, we choose to use the width-at-half-maximum $w^H_i=2.773\,\sigma_i$ to replace $\sigma_i$, then
\begin{equation}
A_i = A'_i\sqrt{2\pi}\,~0.425\,w^H_i,
\end{equation}
where the values of $A'_i$ and $w^H_i$ can be read directly from the plot.
For a rough estimate,
$A'_i\sim 130, 122$ and $95$ and $w^H_i\sim 0.15, 0.3$ and  $0.6$. The estimated $A_i$ values are $21, 39$ and $61$ respectively, 
which are quite consistent with the input values.

Next, we change the setup to a more relevant one: one isolated state ($m_0=0.6$ and $A_0=200$) plus 
a section of dense spectrum (simulating the continuum spectrum in the DIS region) and also some noises.
The dense spectrum is form $\nu=1.5$ to  $\nu=10$ with interval $\delta\nu=0.01$ and weight factor $A=0.5$. 
The signal-to-noise ratio is set to be $100$ at all time slices. In this case, since the noise will cover some details 
of the correlator, we need to use an SVD cutoff of $10^{-6}$ to remove the inaccurate small eigenvalues. 
As show in the right panel of Figure~\ref{fig-2}, the peak shows itself in the right position;
the near constant tail is stable and the value is commendably consistent with the expected value that is $A/\delta\nu=50$.
Again, for the spectrum weight of the isolated state, 
we can read $A'_0\sim280$ and $w^H_0\sim0.7$ from the plot and the estimated $A_0\sim208$ is consistent with the input.

These two toy model tests enable us be confident in the feasibility of applying the Backus-Gilbert method in our hadronic tensor calculation.
We will show the Backus-Gilbert results calculated from our preliminary lattice data in the next section.
Moreover, we will also consider other inverse methods such as the maximum entropy method 
and $\chi^2$ fitting of model spectral function in the future to see whether we can have consistent
results. This will help to estimate the systematic error of the inverse Laplace transform.

\section{Preliminary results}
\hspace{0.5cm}
Our exploratory calculation is carried out on a small $12^3\times 128$ anisotropic clover lattice generated by the 
CLQCD collaboration~\cite{CLQCD_conf} with $m_\pi\sim640$ MeV, 
$a_s=0.1785(53)$ fm and $\xi=5$. The number of configurations used is $500$.

\begin{figure}[tph]
   \centering
   \subfigure[$\vec{p}=0$ case.]%
             {\includegraphics[width=0.475\textwidth,clip]{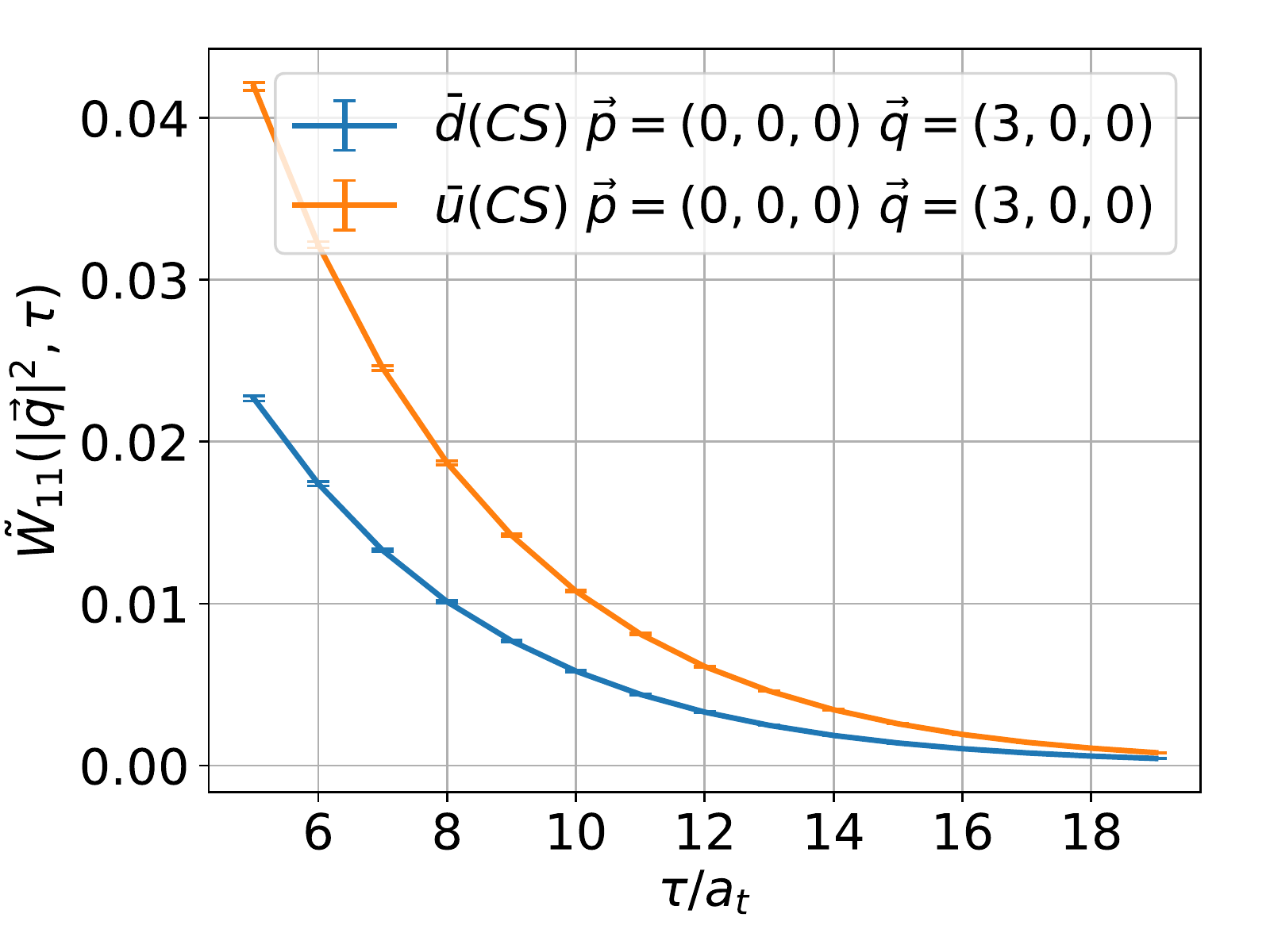}}\hfill
   \subfigure[$\vec{p}\neq0$ case.]%
             {\includegraphics[width=0.475\textwidth,clip]{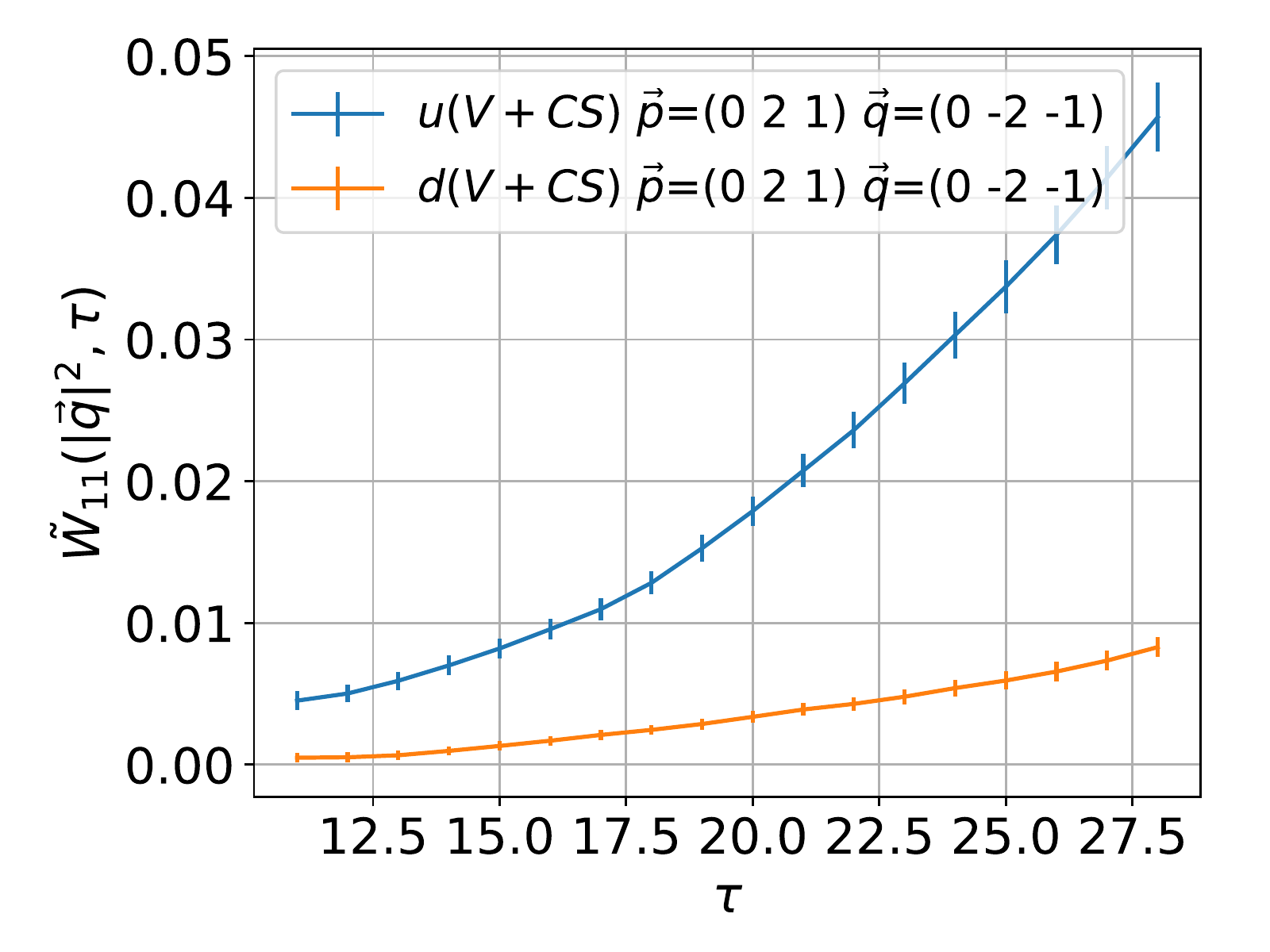}}
   \caption{Euclidean hadronic tensor $\tilde{W}_{11}(\vec{p},\vec{q},\tau)$ as a function of $\tau$.}
   \label{fig-2}
\end{figure}

We choose the direction of the currents $\mu=\nu=1$, so what we calculate is the one-one component of the Euclidean hadronic tensor $\tilde{W}_{11}(\vec{p},\vec{q},\tau)$.
We use two sequential sources (one from $t_0$ through $t_1$ to $t_2$ while the other from $t_0$ through $t_f$ to $t_2$) to calculate the four-point functions. Therefore, 3 times of inversions are needed for each $\vec{p}$, $\vec{q}$, $t_1$ and $t_f$.
The Chroma software \cite{Edwards:2004sx} is used to implement our calculation.
The results of $\tilde{W}_{11}$ are shown in Figure~\ref{fig-2} as a function of the Euclidean time $\tau$. 
The left plot is the connected sea anti-parton case from Fig.~\ref{insertions} (b) with $\vec{p}=0$.
We see that both curves of $\bar{u}$ and $\bar{d}$ go down as $\tau$ increases since the energy of the lowest intermediate state (i.e. nucleon energy
with a momentum of $\vec{p}+ \vec{q}$) is larger than that of the nucleon at rest. The right one is the valence plus connected sea case with 
$\vec{p}\neq0$ and $\vec{q}=-\vec{p}$. Both curves go up as $\tau$ increases since the lowest intermediate state is the nucleon at rest whose energy is lower than that of the initial state with a moving nucleon (i.e. $m_N < E_{\vec{p}}$.)

\begin{figure}[tp]
   \centering
   \subfigure[Minkowski hadronic tensor]%
             {\includegraphics[width=0.475\textwidth,clip]{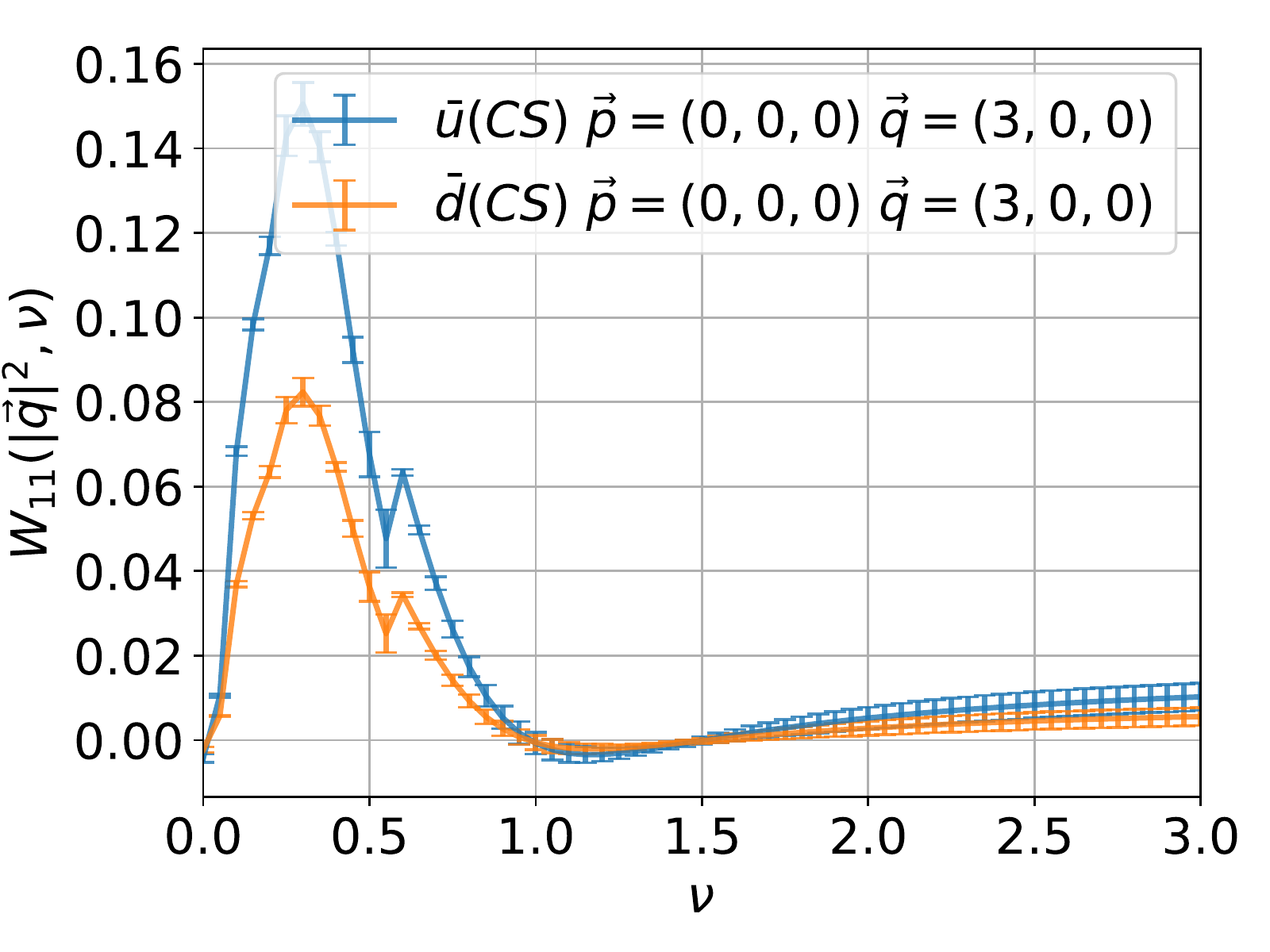}}\hfill
   \subfigure[$\gamma p$ total cross section, from PDG.]%
             {\includegraphics[width=0.475\textwidth,clip]{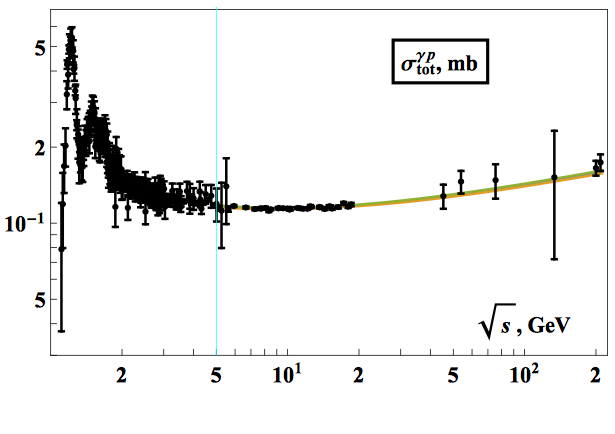}}
   \caption{Minkowski hadronic tensor and the $\gamma p$ total cross section.}
   \label{fig-3}
\end{figure}

The Minkowski hadronic tensor ${W}_{11}(q^2,\nu)$, up to a constant, is just the spectral density $\omega(\nu)$ in the expression of the Backus-Gilbert method. After doing the Backus-Gilbert reconstruction using the data in the $\vec{p}=0$ case,
we get the Minkowski hadronic tensor ${W}_{11}$ as shown in the left plot of Figure~\ref{fig-3}.

The lowest peaks are the elastic peaks at $\nu = E_{\vec{p}+ \vec{q}} - E_{\vec{p}}$ as expected. The second peaks are related to
the quasi-elastic peaks of the $\gamma N$ reaction which are the excited nucleon states which includes Roper state, $S_{11}$, $\pi N$, etc.
When the deep inelastic region sets in as in the experimental $\gamma p$ total cross section $\sigma_{tot}^{\gamma p}$ in the right
panel of Fig.~\ref{fig-3}, the density of the spectral function is expected to be smooth, reflecting the parton nature of DIS. 
For the DIS, the kinematics requires $\nu < |\vec{q}|$ so that $Q^2 > 0$ and $0 \leq x=\frac{Q^2}{2m_p\nu} \leq 1$.
Furthermore, $\nu$ should be several GeV above the quasi-elastic peaks to be in the DIS scaling region. When we examine Fig.~\ref{fig-3} (a),
we find that when the quasi-elastic peak tapers off at $\nu \sim 1$ which is $\sim 5.5$ GeV on this lattice, $|\vec{q}| = 3$ corresponds to
1.7 GeV. Thus, $Q^2 <0$ in this case, there is no room for the physical DIS region. 

\begin{figure}[tp]
   \centering
   \subfigure[]%
             {\includegraphics[width=0.475\textwidth,clip]{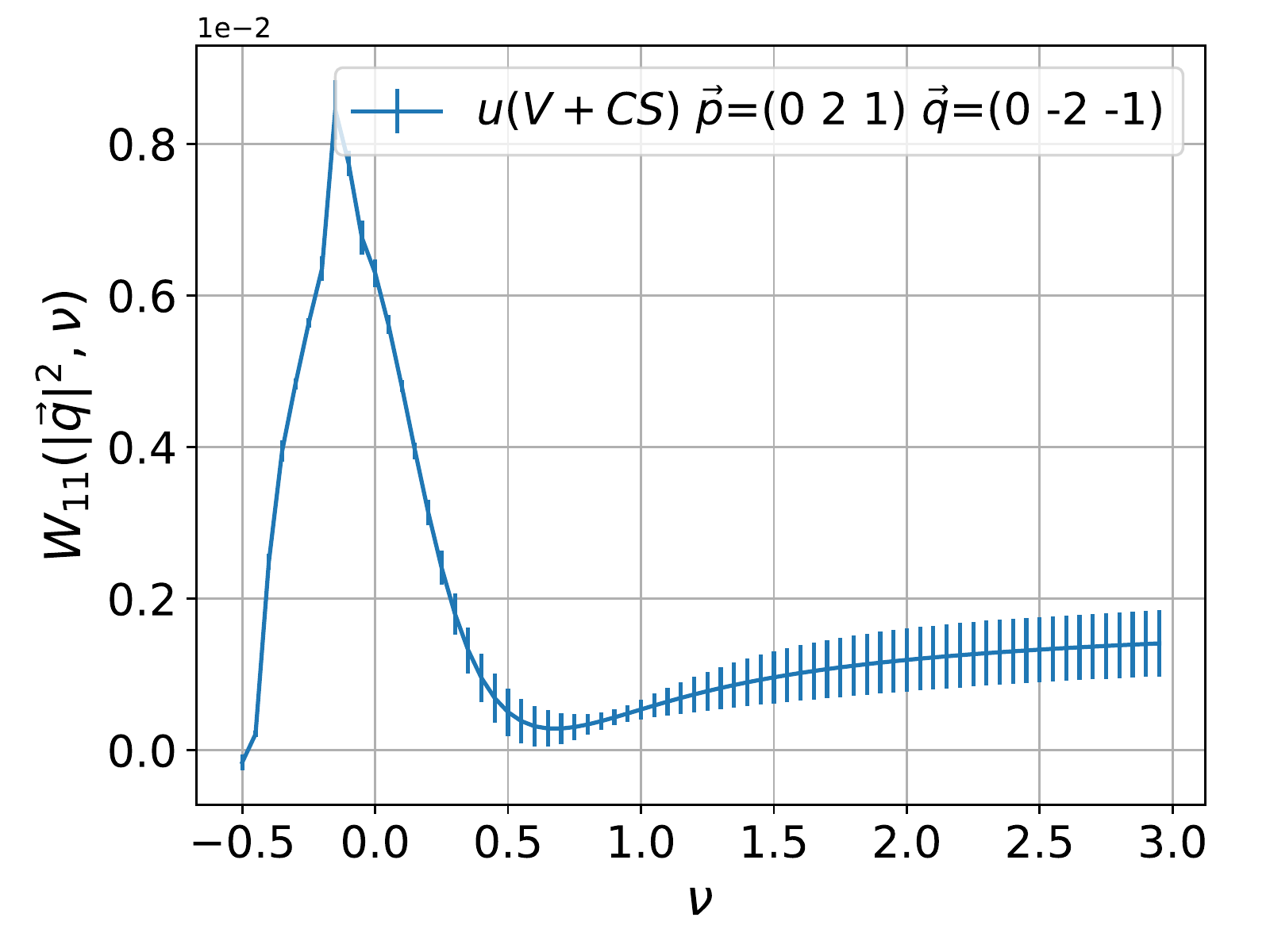}}\hfill
   \subfigure[]%
             {\includegraphics[width=0.475\textwidth,clip]{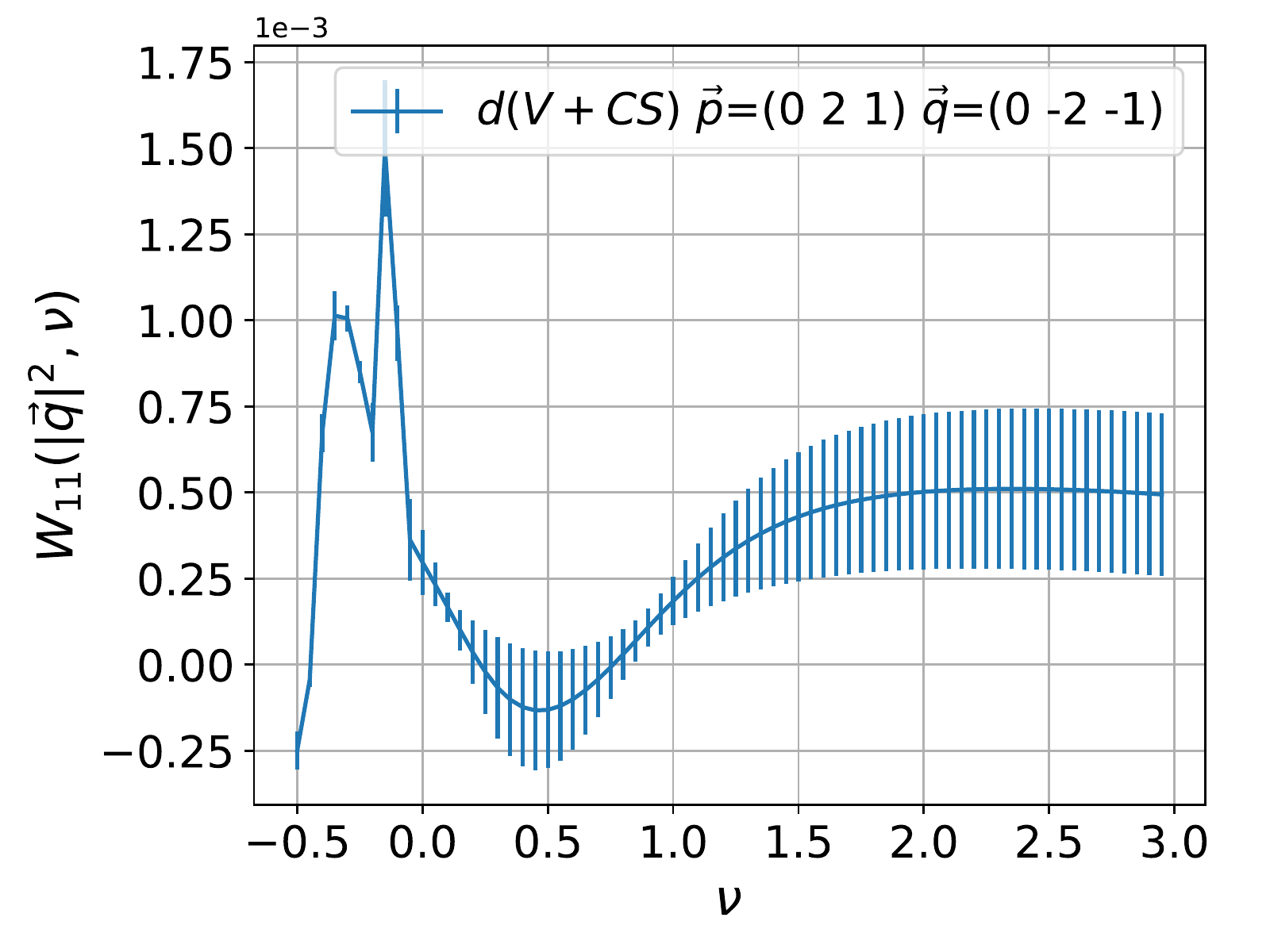}}
   \caption{Minkowski hadronic tensor}
   \label{fig-4}
\end{figure}

We also analyze the $\vec{q}= - \vec{p} \neq 0$ case , the corresponding Euclidean data on $\tilde{W}_{11}(\vec{p},\vec{q},\tau)$ are on the 
right panel of Figure~\ref{fig-2}. The Backus-Gilbert reconstructed results are shown in Figure~\ref{fig-4}.
The elastic peaks  are at $\nu < 0$ since $E_{\vec{p} + \vec{q}} < E_{\vec{p}}$. However, again
the kinematics of the setup still cannot bring us to the physical region of $Q^2$ and $x$, the $Q^2$ at the tail of the inelastic peak at 
$\nu \sim 0.5$ is still negative.

It is clear from these preliminary studies that one needs to have a large $|\vec{q}|$ so that there is enough room for 
the energy transfer $\nu$ to be less than $|\vec{q}|$ and yet a few GeV above the inelastic peak. To achieve this, one needs lattices
with smaller spatial lattice spacings in order to have a relatively large coverage of the range of $x$, e.g. from 0.07 to 0.45.  The fact that both the elastic and quasi-elastic peaks are revealed is, nevertheless,  a positive sign for the 
Backus-Gilbert inverse method.



\bibliography{lattice2017}

\end{document}